\documentclass[10pt,runningheads]{llncs}
\usepackage{booktabs}  % add
\usepackage{multirow}
\usepackage{bm}
\usepackage{array}
\usepackage{graphicx}
\usepackage{epstopdf}
\usepackage{subfigure}
\usepackage{amssymb}
\usepackage{amsmath}
\usepackage{makecell}
\usepackage[pagebackref=false,breaklinks=true,letterpaper=true,colorlinks,bookmarks=false]{hyperref}
\usepackage{cite}
\usepackage[draft,authormarkuptext=none,addedmarkup=none]{changes}
\colorlet{Changes@Color}{blue}
\usepackage{pifont}

\def\eg{{\em e.g.}}

\emergencystretch=\hsize
\graphicspath{{figures/}}

\begin{document}
\title{$\Delta$-Diffusion: 
Modeling Longitudinal Brain Amyloid-PET   Trajectories via Conditional Poisson Diffusion Bridge}

\author{Yongheng~Sun\inst{1}, 
Minhui~Yu\inst{1}, 
Mengqi~Wu\inst{1}, 
Maureen~Kohi\inst{1},
Mingxia~Liu\inst{1,*}}  

\authorrunning{Y.~Sun, M.~Yu, M.~Wu, M.~Kohi, and M.~Liu}

\titlerunning{$\Delta$-Diffusion for 
Longitudinal Brain Amyloid-PET Synthesis}

\institute{Department of Radiology and BRIC, University of North Carolina at Chapel Hill, Chapel Hill, North Carolina 27599, USA\\
$^*$Corresponding author: M.~Liu (\email{mingxia\_liu@med.unc.edu})\\
}
  
\maketitle

\begin{abstract}
While longitudinal brain PET imaging is the gold standard for quantifying the spatiotemporal accumulation of $\beta$-amyloid, its widespread clinical utility is constrained by high operational costs and cumulative radiation risks. 
Recent deep generative models show promise in longitudinal image synthesis; however, they often fail to capture subtle pathological progression 
due to \emph{identity drift} and a persistent bias toward \emph{trivially replicating baseline signal intensities} rather than modeling temporal transition. 
To this end, we propose \textbf{$\Delta$-Diffusion}, a novel progression-aware framework that redefines longitudinal PET synthesis as a conditional Poisson Diffusion Bridge (PDB) process. 
Unlike standard diffusion models that start from Gaussian noise, our PDB formulation is mathematically anchored to the subject's baseline PET, effectively transforming the generative task into a \emph{conditional distribution transition} of the amyloid trajectory. 
To handle heteroscedastic nature of PET imaging, we introduce a physically-grounded Poisson perturbation within a Diffusion Transformer (DiT). 
This architecture uses adaptive scale-shift modulation to precisely calibrate the synthesis with the elapsed clinical interval and structural MRI context. 
A volume-of-interest %aware image constraint 
balanced objective
is  %utilized to focus the model on sparse, high-stakes regions of amyloid accumulation. 
designed to emphasize sparse, high-risk regions of amyloid accumulation. 
Validated on two cohorts with 542 subjects, %ADNI and OASIS3, 
$\Delta$-Diffusion demonstrates superior performance in capturing longitudinal variations in amyloid deposition compared to state-of-the-art methods, offering a robust computational framework for tracking disease progression.

% Authors must provide keywords and are not allowed to remove this Keyword section.

\end{abstract}
\section{Introduction}
Tracking spatiotemporal progression of molecular biomarkers, such as $\beta$-amyloid, is fundamental to understanding the trajectory of neurodegenerative diseases like Alzheimer's~\cite{yang2025hyperbolic,jack2008alzheimer,lamontagne2019oasis}. 
%In clinical practice, 
This is typically quantified using Positron Emission Tomography (PET)  radiotracers such as $^{11}$C-Pittsburgh Compound-B (PIB) and $^{18}$F-Florbetapir (AV45)~\cite{landau2013amyloid,hsiao2012correlation}. 
While longitudinal PET serves as the gold standard for monitoring these changes, the acquisition of serial scans is frequently hindered by cumulative radiation risks, high costs, and the high rate of subject attrition in long-term studies~\cite{lawrence2017systematic,furuse2019radiological}. There is a critical need for computational methods capable of forecasting future PET states from a single baseline session, providing a non-invasive way to impute follow-up data or predict disease staging.

Despite advances in deep learning-based medical image synthesis~\cite{yu2026rela}, static PET synthesis models remain inadequate for longitudinal forecasting.
\emph{First}, because follow-up scans are anchored to the same subject~\cite{sun2023dual, sun2026dumeta++}, models often suffer from spatial-identity drift~\cite{huang2025identity} or, conversely, a tendency to trivially copy the baseline scan without capturing progression~\cite{gong2024individualised}. 
\emph{Second}, clinical follow-up intervals are highly heterogeneous; ignoring the elapsed time leads to temporally miscalibrated results. 
\emph{Third}, the signal-dependent nature of PET noise (governed by Poisson statistics) is poorly modeled by Gaussian diffusion. This necessitates a physically-grounded model to preserve critical signals in low-SNR regions inherent to PET's photon-counting physics. 

\begin{figure}[!t]
\centering
\includegraphics[width=\textwidth]{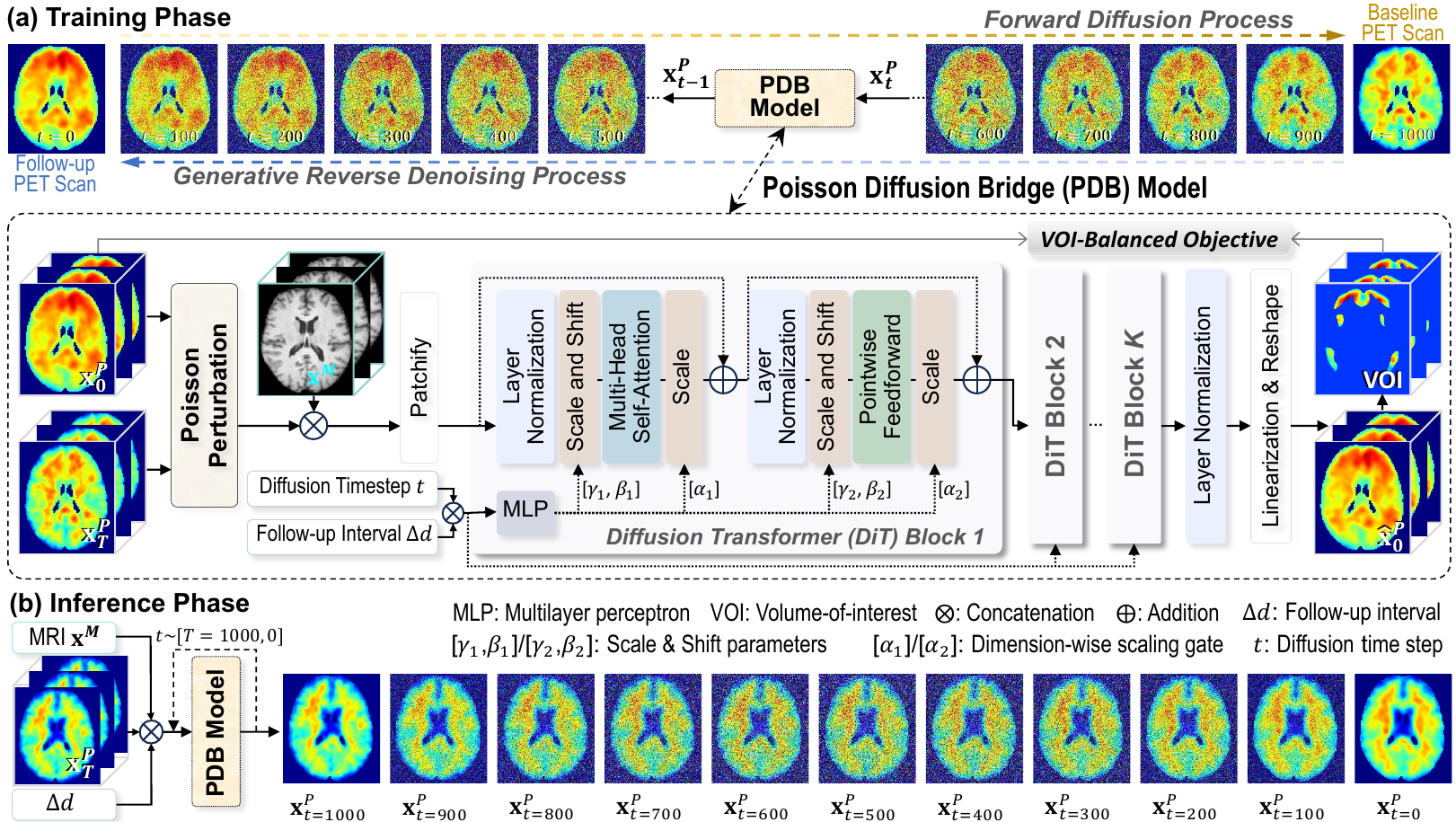}
\caption{{Overview of 
$\Delta$-Diffusion.} 
\textbf{(a) Training Phase:} 
The framework learns a transition score for a {Poisson Diffusion Bridge} (PDB) process mapping baseline PET $\mathbf{x}_T^P$ to follow-up PET $\mathbf{x}_0^P$. By anchoring the bridge to baseline, DiT blocks explicitly model the distributional trajectory of pathological progression ($\Delta$), guided by structural T1w MRI $\mathbf{x}^M$ and the clinical interval $\Delta d$ under a VOI-balanced objective. \textbf{(b) Inference Phase:} A follow-up PET is synthesized by solving the reverse-time bridge SDE starting from the observed baseline $\mathbf{x}_T^P$, ensuring anatomical and temporal consistency.
} 
\label{fig_pipeline}
\end{figure}

To address these limitations, we propose a progression-aware \textbf{$\Delta$-Diffusion} framework. 
Unlike traditional generative models that treat the baseline as a secondary condition, our method reformulates longitudinal synthesis as a Poisson Diffusion Bridge (PDB) process. 
As shown in Fig.~\ref{fig_pipeline}(a), we construct a diffusion bridge that is mathematically anchored to the subject's baseline PET.
This formulation redefines the generative objective as learning the \emph{conditional distribution transition}, effectively mitigating the trivial identity task and ensuring strict subject-specific consistency. 
To improve physiological fidelity, we incorporate a physically-grounded Poisson perturbation layer and a Diffusion Transformer (DiT)  that explicitly calibrates the output based on the follow-up interval and anatomical context from T1-weighted (T1w) MRI. 
We also design a volume-of-interest (VOI)-balanced objective to prioritize learning in clinically significant regions. 
Experiments on two cohorts ($N$=542) 
with longitudinal PIB- and AV45- PET scans demonstrate that our method significantly outperforms state-of-the-art (SOTA) approaches 
in predicting 
longitudinal amyloid variations.

\section{Proposed Method}

\subsubsection{Problem Formulation.} 
Let $\mathcal{X}^P$ denote the domain of amyloid PET scans and $\mathcal{X}^M$ denote the domain of structural T1w MRI scans. 
Given a subject's PET scan $\mathbf{x}_T^P \in \mathcal{X}^P$ acquired at {the baseline time} $d_1$, our goal is to predict the follow-up PET scan $\mathbf{x}_0^P \in \mathcal{X}^P$ at a follow-up {time point} $d_2 = d_1 + \Delta d$. 
This task is formulated as learning a conditional transition distribution:
\begin{equation}
\mathbf{x}_0^P \sim p_\theta(\mathbf{x}_0^P \mid \mathbf{x}_T^P, \mathbf{x}^M, \Delta d),
\end{equation}
where the generation is conditioned on the T1w MRI $\mathbf{x}^M \in \mathcal{X}^M$ and the temporal interval $\Delta d$. 
We frame this longitudinal synthesis as a \emph{conditional diffusion bridge} process. 
Unlike standard diffusion models that map unstructured Gaussian noise to the data distribution~\cite{ho2020denoising,songdenoising}, we learn a stochastic trajectory $(\mathbf{x}_t)_{t \in [0,T]}$ anchored to subject-specific baseline at the endpoint $T$ (i.e., $\mathbf{x}_T = \mathbf{x}_T^P$) and relaxing toward the follow-up state at $t=0$ according to the elapsed interval $\Delta d$. 
During inference, as illustrated in Fig.~\ref{fig_pipeline}(b), the follow-up PET is synthesized by solving the reverse-time bridge dynamics starting from the observed baseline state.

\subsection{Diffusion Bridge for Longitudinal Synthesis}
\paragraph{\textbf{Diffusion Bridge.}}
Following~\cite{zhou2023denoising}, we consider a continuous-time diffusion process $\{\mathbf{x}_t\}_{t\in[0,T]}$ governed by a Stochastic Differential Equation (SDE): 
\begin{equation}
\small 
d\mathbf{x}_t = {\mathbf{f}}(\mathbf{x}_t,t)\,dt + g(t)\,d\mathbf{w}_t,
\end{equation}
{where ${\mathbf{f}}(\cdot,\cdot)$ and $g(\cdot)$ denote the drift and diffusion terms,
respectively, and $\mathbf{w}_t$ is a standard Wiener process.
We} construct a \emph{bridge} that is conditioned to hit a prescribed endpoint at $t=T$. Concretely, letting the endpoint be $\mathbf{x}_T=\mathbf{y}$, Doob’s $h$-transform yields the drift-adjusted forward bridge SDE:
\begin{equation}
\small 
d\mathbf{x}_t
= {\mathbf{f}}(\mathbf{x}_t,t)\,dt + g(t)^2\,{\mathbf{h}}(\mathbf{x}_t,t,\mathbf{y},T)\,dt + g(t)\,d\mathbf{w}_t,
%\quad 
% ~~{\mathbf{x}_0=\mathbf{x},}
~~\mathbf{x}_T=\mathbf{y},
\end{equation}
where {\small{
$\mathbf{h}(\mathbf{x}_t,t,\mathbf{y},T)
= \nabla_{\mathbf{x}_t}\log p(\mathbf{x}_T\mid \mathbf{x}_t)\big|_{\mathbf{x}_T=\mathbf{y}}$}} 
%\end{equation}
is the score function (log-gradient) %gradient of the log transition density 
of the unconditioned diffusion from $t$ to $T$. 
In longitudinal settings, we anchor the bridge endpoint to the observed baseline PET scan, defining: 
%In longitudinal setting, we pin the endpoint to the observed baseline PET and define
\begin{equation}
\small 
\mathbf{x}_T \equiv \mathbf{x}_T^{P} \;\;(\text{baseline PET}), ~~
\mathbf{x}_0 \equiv \mathbf{x}_0^{P} \;\;(\text{follow-up PET}).
\end{equation}
Following this, longitudinal synthesis is achieved by integrating the reverse-time bridge SDE from  {\small{$t=T$}} back to  {\small{$t=0$}}. 
By initializing the solver at the observed baseline {$\mathbf{x}_T^P$}, the model generates a sample distributed as the follow-up PET.

\paragraph{\textbf{Reverse-Time Bridge Dynamics.}}
The reverse-time SDE for sampling the conditional bridge can be formulated as:
\begin{equation}
\small 
d\mathbf{x}_t
=
\Big[{\mathbf{f}}(\mathbf{x}_t,t) - g(t)^2\big({\mathbf{s}}(\mathbf{x}_t,t,\mathbf{y},T)-{\mathbf{h}}(\mathbf{x}_t,t,\mathbf{y},T)\big)\Big]dt
+ g(t)\,d\bar{\mathbf{w}}_t,
~~ \mathbf{x}_T=\mathbf{y},
\end{equation}
where ${\mathbf{s}}(\mathbf{x}_t,t,\mathbf{y},T)$$=$$\nabla_{\mathbf{x}_t}\log q(\mathbf{x}_t\mid \mathbf{x}_T$$=$$\mathbf{y})$ is the (intractable) conditional score of intermediate states given the endpoint. In practice, we learn a neural score network $s_{\theta}$ and replace $s$ with $s_{\theta}$ in the reverse solver to generate $\mathbf{x}_0^{P}$ from $\mathbf{x}_T^{P}$.
For inference-time sampling, we directly adopt a higher-order hybrid solver~\cite{zhou2023denoising}, which builds on a second-order Heun ODE sampler and interleaves scheduled Euler--Maruyama steps to retain stochasticity and avoid overly averaged outputs.

\paragraph{\textbf{Conditioning.}} 
We parameterize $s_{\theta}$ with a DiT (comprising $K$=28 blocks, following~\cite{li2025diffusion}). 
{To incorporate temporal guidance}, we inject \textbf{(i)} diffusion time step $t$ and \textbf{(ii)} follow-up interval $\Delta d$ (in days) through multilayer perceptron (MLP) embeddings. % that predict adaptive scale-shift and gating parameters within each DiT block.  
These embeddings predict adaptive scale–shift and gating parameters within each DiT block, effectively modulating the generative trajectory based on the elapsed time between scans. 
This design is aligned with adaptive layer normalization used in volumetric transformers. 
{For structural guidance}, the baseline T1w MRI $\mathbf{x}^{M}$ is integrated as a conditional input to provide anatomy priors,  while the bridge endpoint $\mathbf{x}_T^P$ anchors the subject-specific baseline. 

\enlargethispage*{0.8\baselineskip}
\subsection{Physically-Grounded Poisson Perturbation} 
Standard diffusion/bridge constructions often rely on Gaussian perturbations. However, PET noise is dominated by Poisson counting statistics~\cite{sanaat2020projection}. 
%Motivated by this, 
To address this physical characteristic, we replace the forward perturbation with a Poisson corruption operator. 
Let $\mathbf{z}_t$ denote the perturbed sample at diffusion step $t\in\{0,\cdots,T\}$. % in DDBM. 
Mathematically, Poisson diffusion can be written as:
\begin{equation}
\small 
\mathbf{z}_t = \mathrm{perturb}(\mathbf{z}_{t-1},\lambda_t),
~~ \lambda_1 < \lambda_2 < \cdots < \lambda_T,
\end{equation}
where $\lambda_t$ controls the noise intensity and is typically increased with $t$ (e.g., linearly interpolated in $[0,1]$). 
{Physically-grounded Poisson perturbation can be implemented by applying Poisson deviates in the projection (sinogram) domain~\cite{jiang2023pet}. 
The sinogram is generated via a Radon transform to emulate acquisition-related photon noise, followed by an inverse reconstruction to obtain $\mathbf{z}_t$.}

Although image-domain noise from Poisson sinogram perturbations is not strictly 
i.i.d. Gaussian, it can be approximated as heteroscedastic Gaussian under moderate-to-high count regimes~\cite{fessler1994penalized}.  
This justifies adopting the standard scheduling %of Denoising diffusion bridge model
used in~\cite{zhou2023denoising} as a practical approximation, while keeping the forward process physically grounded in Poisson counting statistics. 
In our framework, Poisson perturbation is applied in the forward process to generate realistic intermediate states and enhance the robustness of the reverse denoising trajectory. 

\subsection{VOI-Balanced Objective} 
Given that amyloid burden is clinically evaluated in specific cortical regions,  we propose a VOI-balanced objective using a cortical VOI {$\mathbf{V}\in\{0,1\}^{\Omega}$} from~\cite{klunk2015centiloid} to prioritize clinically significant regions for amyloid quantification while maintaining global structural integrity. 
Let $\mathbf{x}$ and $\hat{\mathbf{x}}$ denote ground-truth and predicted images. 
We compute the region-normalized mean squared error (MSE) as: 
{
\begin{equation}
%\small
\footnotesize
%\scriptsize
\mathrm{MSE}(\hat{\mathbf{x}},\mathbf{x};\mathbf{V})
=
\rho \frac{\sum_{{i}\in\Omega}\mathbf{V}({i})\left(\hat{\mathbf{x}}({i})-\mathbf{x}({i})\right)^2}
{\sum_{{i}\in\Omega}\mathbf{V}({i})+\varepsilon} 
+
(1-\rho)\frac{\sum_{i\in\Omega}\mathbf{(1-V)}({i})\left(\hat{\mathbf{x}}({i})-\mathbf{x}({i})\right)^2}
{\sum_{{i}\in\Omega}\mathbf{(1-V)}({i})+\varepsilon},
\end{equation}
where $\varepsilon$ is a small constant for numerical stability, and $\rho$ (empirically set as $0.8$)} controls the contributions of VOI-masked regions and the background.
The objective function at the diffusion time step $t$ of our framework is defined as: 
\begin{equation}
\mathcal{L}(t)=w_t \, \mathrm{MSE}(\hat{\mathbf{x}},\mathbf{x};\mathbf{V})
%\mathcal{L}_{\mathrm{balMSE}},
\end{equation}
where $w_t$ is a predefined weight determined by bridge schedule parameters~\cite{zhou2023denoising}.

\begin{table}[!t]
 \renewcommand{\arraystretch}{1}
%\small
%\footnotesize
\scriptsize
\centering
\setlength\tabcolsep{0.3pt}
\caption{Quantitative results achieved by $\Delta$-Diffusion and eight competing methods for longitudinal PIB-PET generation on {OASIS-3}~\cite{lamontagne2019oasis}, with best results shown in bold.}
%\label{tab:comp_pib}
%\scalebox{1}{
\resizebox{1\textwidth}{!}{
\begin{tabular}{@{}l|cc|cc|c@{}}
\toprule
\multicolumn{1}{l|}{\multirow{2}{*}{{~Method}}} 
& \multicolumn{2}{c|}{{Whole-Image Evaluation}}
& \multicolumn{2}{c|}{{Region-Level Evaluation}}
& \multicolumn{1}{c}{\multirow{2}{*}{ $p$-value }} 
\\
\cline{2-5}
& PSNR    & SSIM   & PSNR & SSIM  & %$p$-value 
\\ 
\midrule
~VAEGAN\cite{larsen2016autoencoding}           & 27.1544$_{\pm1.8800}$ & 0.8692$_{\pm0.0208}$ & 36.5221$_{\pm3.4173}$           & 0.9812$_{\pm0.0042}$ &$4.92$$\times$$10^{-33}$                        \\
~DDPM\cite{ho2020denoising}          & 24.3193$_{\pm2.0479}$ & 0.8286$_{\pm0.0301}$ & 35.5428$_{\pm3.4597}$           & 0.9790$_{\pm0.0064}$&$4.94$$\times$$10^{-56}$                       \\
~DDIM\cite{songdenoising}          & 21.1245$_{\pm1.1318}$ & 0.7563$_{\pm0.0137}$ & 31.6217$_{\pm2.1487}$           & 0.9742$_{\pm0.0050}$&$2.64$$\times$$10^{-87}$                       \\
~PNDM\cite{liupseudo}          & 11.4500$_{\pm0.1907}$ & 0.7340$_{\pm0.0111}$ & 20.5016$_{\pm0.3231}$           & 0.9619$_{\pm0.0017}$&$1.97$$\times$$10^{-132}$                       \\
~LDM\cite{rombach2022high}           & 25.0878$_{\pm2.5740}$ & 0.8487$_{\pm0.0301}$ & 35.4552$_{\pm3.6322}$           & 0.9792$_{\pm0.0053}$&$1.40$$\times$$10^{-51}$                       \\
% VQVAE         & 30.0691$_{\pm3.7635}$ & 0.9123$_{\pm0.0316}$ & 38.9555$_{\pm3.5908}$           & 0.9597$_{\pm0.0068}$                       \\
~ControlNet\cite{zhang2023adding}    & 24.3533$_{\pm2.1836}$ & 0.8133$_{\pm0.0387}$ & 34.7076$_{\pm2.1922}$           & 0.9777$_{\pm0.0080}$&$7.19$$\times$$10^{-32}$                       \\
~FICD\cite{yu2024functional}          & 24.2175$_{\pm1.6808}$ & {0.8297}$_{\pm0.0343}$ & 33.0675$_{\pm2.4198}$           & 0.9761$_{\pm0.0090}$&$3.50$$\times$$10^{-36}$                       \\
~SiM2P\cite{li2025diffusion}  & 25.1127$_{\pm2.3921}$        & 0.8200$_{\pm0.0242}$       & 34.4062$_{\pm3.0458}$        & 0.9777$_{\pm0.0060}$&$3.67$$\times$$10^{-60}$ 
\\                       
\midrule
%\hline
~$\Delta$-Diffusion          & \textbf{28.6489}$_{\pm1.9933}$ & \textbf{0.8732}$_{\pm0.0232}$ & \textbf{38.0669}$_{\pm3.5655}$ & \textbf{0.9841}$_{\pm0.0055}$ &--                      \\ \bottomrule
\end{tabular}
}
\label{tab_OASIS}
\end{table}

\enlargethispage*{0.8\baselineskip}
\section{Experiment}
\paragraph{\textbf{Materials and Image Preprocessing.}} 
Two datasets are used. 
(1) \textbf{{OASIS-3}}~\cite{lamontagne2019oasis}: 
283 subjects have paired baseline PIB-PET and T1w MRI, {with 1--4 follow-up PIB-PET sessions}, yielding 653 longitudinal pairs. 
{These samples were split into training/validation (463/51 pairs, 226 subjects) and test (139 pairs, 57 subjects) sets.} 
(2) \textbf{ADNI}~\cite{jack2008alzheimer}:  
259 subjects have paired baseline AV45-PET and T1w MRI, {with 1--5 follow-up AV45-PET sessions}, yielding 835 longitudinal pairs. 
{These samples were split into training/validation (614/68 pairs, 207 subjects) and test (153 pairs, 52 subjects) sets.} 
Data splits were performed at the \emph{subject level} to avoid data leakage, ensuring that no subject appears in more than one set. 
For MRIs, the preprocessing pipeline includes skull stripping, intensity normalization, and non-linear registration to the MNI space. 
PET scans are first skull-stripped and linearly aligned to their respective MRIs; they are then warped to the MNI space using MRI-derived deformation fields. 
To remove non-informative background voxels, 
all volumes are uniformly cropped to {dimensions} of $160$$\times$$ 180$$\times$$160$. 
To facilitate inter-subject comparisons of tracer uptake, standardized uptake value ratios (SUVRs) are computed for each subject by normalizing regional tracer uptake using the \emph{cerebellum} as reference region~\cite{klunk2015centiloid}.

\begin{table}[!t]
 \renewcommand{\arraystretch}{0.9}
%\footnotesize
\scriptsize
\centering
\setlength\tabcolsep{0.3pt}
\caption{Quantitative results achieved by $\Delta$-Diffusion and eight competing methods for longitudinal 
AV45-PET generation on ADNI~\cite{jack2008alzheimer}, with best results shown in bold.}
%\label{tab:comp_av45}
%\scalebox{1}{
\resizebox{1\textwidth}{!}{
\begin{tabular}{@{}l|cc|cc|c@{}}
\toprule
\multicolumn{1}{l|}{\multirow{2}{*}{~{Method}}} 
& \multicolumn{2}{c|}{{Whole-Image Evaluation}}
& \multicolumn{2}{c|}{{Region-Level Evaluation}}
& \multicolumn{1}{c}{\multirow{2}{*}{ $p$-value }} 
\\
\cline{2-5}
& PSNR    & SSIM   & PSNR & SSIM  & %$p$-value 
\\ \midrule
~VAEGAN\cite{larsen2016autoencoding}           & 23.9580$_{\pm2.1268}$ & 0.8148$_{\pm0.0273}$ & {33.3028}$_{\pm2.6115}$           & 0.9798$_{\pm0.0056}$  & $1.68$$\times$$10^{-17}$                     \\
~DDPM\cite{ho2020denoising}          & 20.1783$_{\pm1.4348}$ & 0.7547$_{\pm0.0094}$ & 31.1558$_{\pm2.8927}$           & 0.9743$_{\pm0.0073}$ &$5.98$$\times$$10^{-49}$                      \\
~DDIM\cite{songdenoising}          & 19.0182$_{\pm1.3541}$ & 0.7391$_{\pm0.0091}$ & 29.6995$_{\pm2.3768}$           & 0.9718$_{\pm0.0061}$  &$1.37$$\times$$10^{-59}$                     \\
~PNDM\cite{liupseudo}          & 11.1641$_{\pm0.3076}$ & 0.7295$_{\pm0.0044}$ & 20.2829$_{\pm0.4048}$           & 0.9600$_{\pm0.0027}$   &$1.41$$\times$$10^{-113}$                     \\
~LDM\cite{rombach2022high}           & 22.0798$_{\pm2.8185}$ & 0.7950$_{\pm0.0337}$ & 31.0900$_{\pm3.6755}$           & 0.9756$_{\pm0.0104}$   &$5.29$$\times$$10^{-20}$                     \\
% VQVAE         & 26.2378$_{\pm3.1325}$ & 0.8577$_{\pm0.0457}$ & 35.1195$_{\pm3.3115}$           & 0.9564$_{\pm0.0084}$                       \\
~ControlNet\cite{zhang2023adding}    & 18.3631$_{\pm2.1888}$ & 0.7297$_{\pm0.0185}$ & 28.1682$_{\pm3.2743}$           & 0.9652$_{\pm0.0085}$  &$5.83$$\times$$10^{-59}$                      \\
~FICD\cite{yu2024functional}          & 23.7418$_{\pm2.2595}$ & 0.8135$_{\pm0.0272}$ & 32.6071$_{\pm2.8574}$           & 0.9526$_{\pm0.0065}$ &$1.04$$\times$$10^{-14}$                       \\
~SiM2P\cite{li2025diffusion}  & 24.5516$_{\pm2.2117}$ & 0.8152$_{\pm0.0237}$ & 33.4698$_{\pm2.6354}$           & 0.9794$_{\pm0.0060}$  &$3.18$$\times$$10^{-25}$                     \\
\midrule
~$\Delta$-Diffusion         & \textbf{25.1723}$_{\pm2.6431}$ & \textbf{0.8284}$_{\pm0.0309}$ & \textbf{34.2089}$_{\pm3.1333}$           & \textbf{0.9806}$_{\pm0.0066}$ &  --                       \\ \bottomrule
\end{tabular}
}
\label{tab_ADNI}
\end{table}
 
\paragraph{\textbf{Experimental Settings.}}
We evaluate PET generation using both \emph{whole-image} and \emph{region-level} metrics: PSNR and SSIM over the full 3D volume for measuring global reconstruction quality, and within the cortical VOI~\cite{klunk2015centiloid} to measure accuracy of localized tracer uptake.  
We compare $\Delta$-Diffusion with SOTA generative 
methods: a hybrid adversarial model (\textbf{VAEGAN}~\cite{larsen2016autoencoding}), three standard diffusion frameworks (\textbf{DDPM}~\cite{ho2020denoising}, \textbf{DDIM}~\cite{songdenoising}, and \textbf{PNDM}~\cite{liupseudo}), and a latent diffusion model (\textbf{LDM}~\cite{rombach2022high}). 
{We also compare our method with \textbf{ControlNet}~\cite{zhang2023adding} for structure-conditioned synthesis, as well as \textbf{FICD}~\cite{yu2024functional} and \textbf{SiM2P}~\cite{li2025diffusion} for medical image translation.}
For all the competing methods, we utilize the same training/validation/test splits and preprocessing pipeline to ensure a fair comparison. 
{All methods are trained on a cluster with NVIDIA H100 GPUs, 
using the Adam optimizer (learning rate: 1$\times 10^{-4}$, batch size: 1).}

\enlargethispage*{0.8\baselineskip}
\paragraph{\textbf{Quantitative Results.}}
As shown in Table~\ref{tab_OASIS} and Table~\ref{tab_ADNI}, in terms of \emph{global reconstruction} quality, our $\Delta$-Diffusion improves PSNR by approximately 3.54 dB on {OASIS-3} (28.6489 vs. 25.1127 for SiM2P) and 0.62 dB on ADNI, demonstrating superior anatomical fidelity. Furthermore, our method achieves the highest accuracy in \emph{localized tracer uptake}, as evidenced by the superior metrics recorded within VOIs. 
While advanced diffusion-based models like LDM and SiM2P generally offer improved stability over standard voxel-space diffusion (\eg, DDPM), they still exhibit significantly lower structural similarity and higher intensity errors compared to our framework. 
Paired $t$-tests on whole-image PSNR show that $\Delta$-Diffusion significantly outperforms the SOTA methods ($p<0.05$).

\begin{figure*}[!t]
\centering
\includegraphics[width=0.98\textwidth]{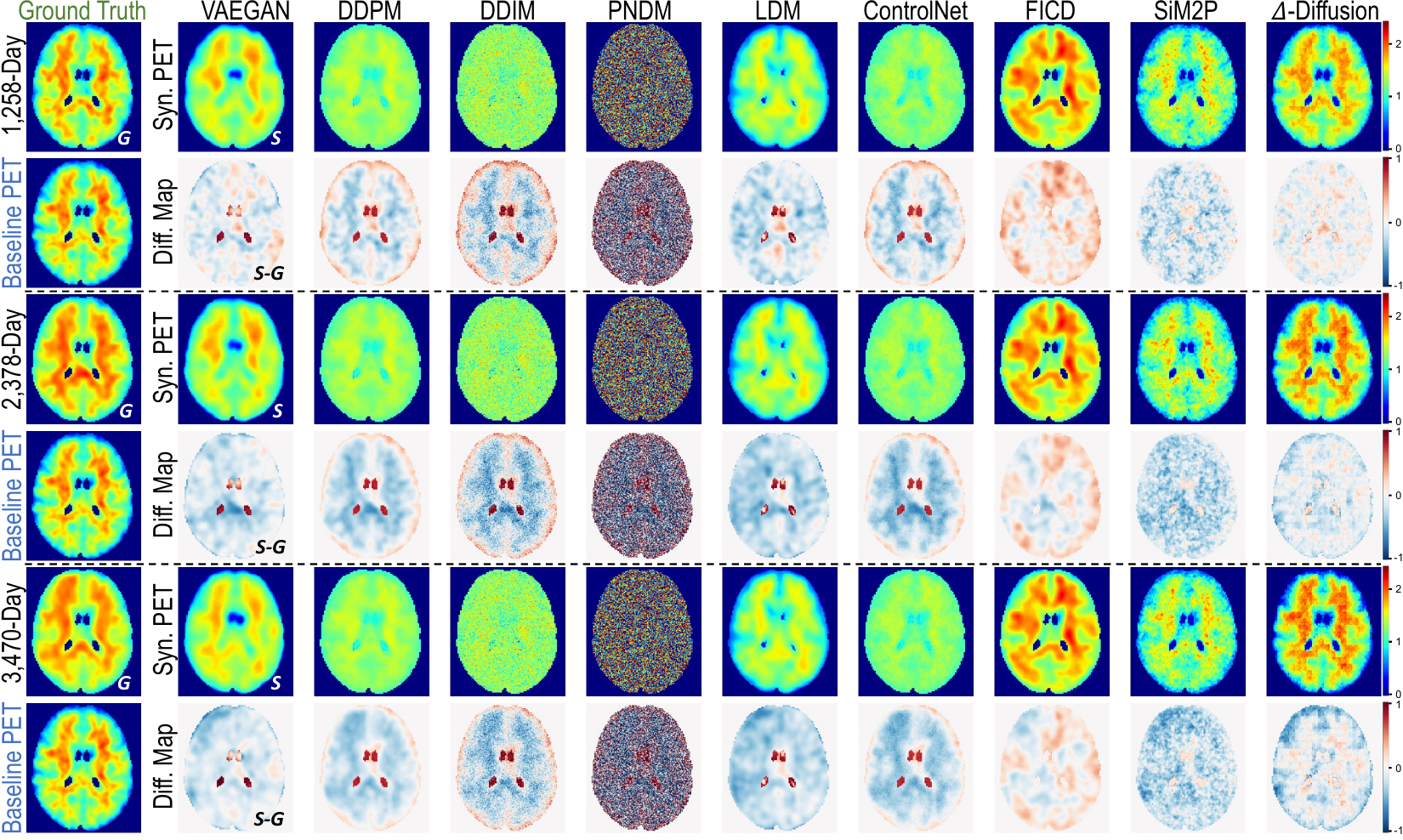}
\caption{Visual comparison of longitudinal PIB-PET  synthesis results on OASIS-3~\cite{lamontagne2019oasis}.} 
\label{fig_visualPIB_OASIS}
\end{figure*}

\begin{figure*}[!t]
\centering
\includegraphics[width=0.98\textwidth]{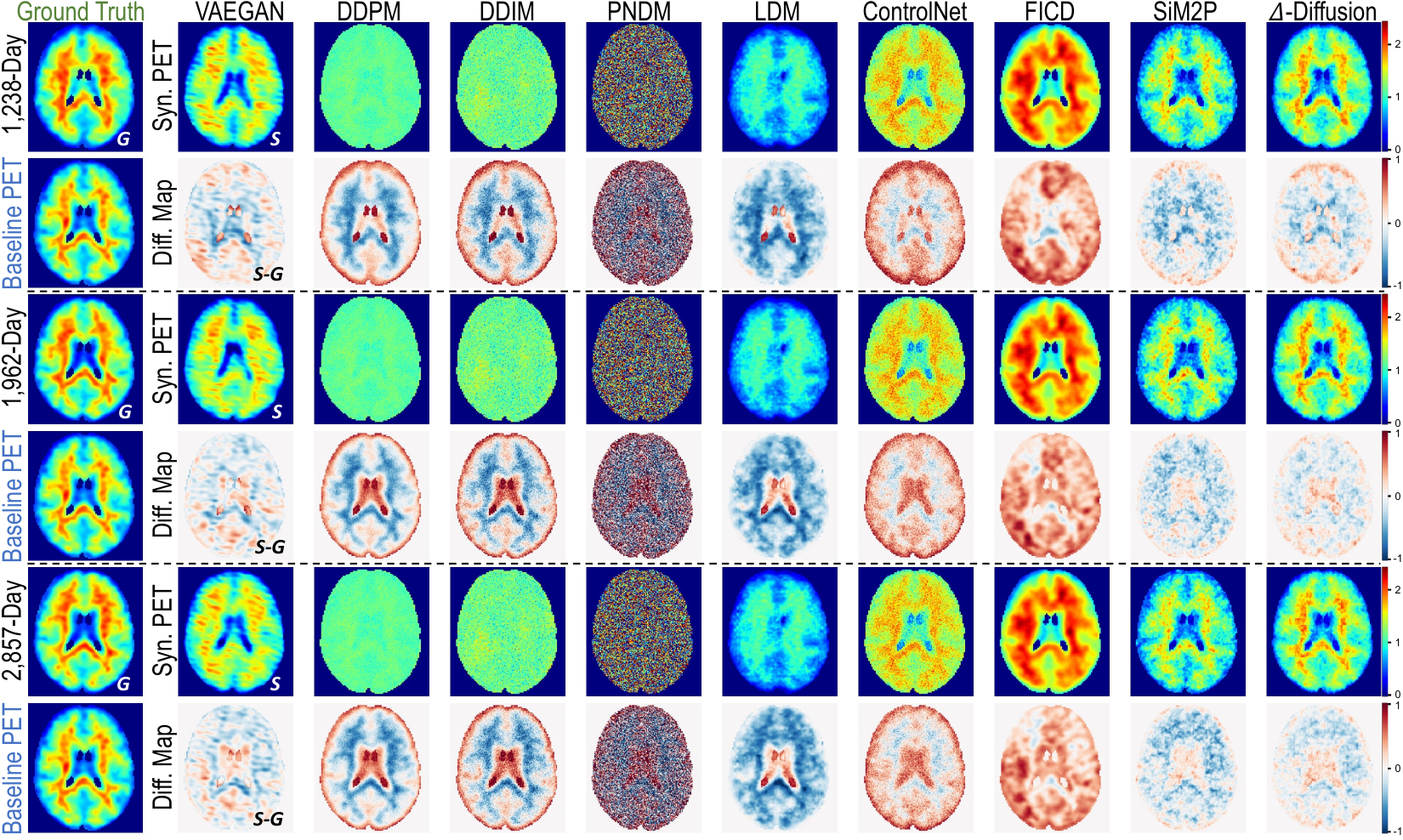}
\caption{Visual comparison of longitudinal AV45-PET synthesis results on ADNI~\cite{jack2008alzheimer}.
} 
\label{fig_visualAV45_ADNI}
\end{figure*}

\paragraph{\textbf{Qualitative Results.}} 
The difference (Diff.) maps in Figure~\ref{fig_visualPIB_OASIS} and Figure~\ref{fig_visualAV45_ADNI} reveal that $\Delta$-Diffusion achieves the lowest reconstruction error, maintaining high fidelity to the baseline scans while capturing longitudinal tracer progression. 
VAEGAN produces visually plausible images but shows significant residuals in the difference maps, suggesting a failure to capture subject-specific functional changes. 
DDPM, DDIM, and PNDM exhibit the most severe intensity discrepancies, with the difference maps showing large-scale errors in cortical regions. 
While LDM, ControlNet, FICD, and SiM2P offer better structural alignment, they still struggle with localized precision, often underestimating or overestimating uptake as evidenced by the residuals. 
In particular, our $\Delta$-Diffusion effectively minimizes these errors across all time points (\eg, up to {3470 days in OASIS-3}), demonstrating its superior ability to model long-term tracer kinetics.

\paragraph{\textbf{Ablation Study.}} 
We conduct an ablation study on $\Delta$-Diffusion using {OASIS-3} data, evaluating performance 
under five settings: 
(1) 
\textbf{w/o~PMVD}: the baseline configuration using Gaussian noise (instead of Poisson), removing all conditioning inputs (MRI and $\Delta d$) and {replacing the VOI-balanced objective with a whole-image MSE loss;  
(2) 
\textbf{w/o~MVD}: using Poisson noise and the whole-image MSE loss, but excluding MRI and $\Delta d$;  
(3) 
\textbf{w/o~MD}: removing MRI and $\Delta d$; 
(4) 
\textbf{w/o~VD}: using the whole-image MSE loss and removing $\Delta d$; 
(5) 
\textbf{w/o~MV}: removing MRI and using the whole-image MSE loss.}  
As shown in Table~\ref{tab_ablation}, removing PMVD causes the largest PSNR drop (28.6489 to 25.1127), confirming the importance of Poisson perturbation and structural-temporal guidance.
Comparing w/o MV to the full model shows that the inclusion of structural MRI and VOI-balanced objective further refines global structural similarity and localized region-level accuracy. 
Even with reduced conditioning, the model maintains high structural fidelity; however, fully integrating temporal and spatial priors in $\Delta$-Diffusion is crucial for achieving the lowest errors in clinical tracer-specific VOIs.

\begin{table}[!t]
 \renewcommand{\arraystretch}{0.8}
\scriptsize
\caption{Ablation results of the proposed $\Delta$-Diffusion for longitudinal PIB-PET generation on {OASIS-3}~\cite{lamontagne2019oasis}, with best results shown in bold.}
\setlength\tabcolsep{0.3pt}
\resizebox{1\textwidth}{!}{
%\scalebox{0.8}{
\begin{tabular}{l|cc|cc}
\toprule
\multirow{2}{*}{Method}             & \multicolumn{2}{c|}{Whole-Image Evaluation}  & \multicolumn{2}{c}{Region-Level Evaluation}  \\ \cline{2-5} 
                                    & PSNR                  & SSIM                 & PSNR                  & SSIM                 \\ 
\midrule
%only Bridge (P: Possion, M: MRI; V: VOI mask; T: timepoint)
$\Delta$-Diffusion~w/o~PMVD & 25.1127$_{\pm2.3921}$ & 0.8200$_{\pm0.0242}$ & 34.4062$_{\pm3.0458}$ & 0.9777$_{\pm0.0060}$ \\
%Bridge+Poisson    
$\Delta$-Diffusion~w/o~MVD& 27.7430$_{\pm2.0588}$ & 0.8558$_{\pm0.0239}$ & 36.7672$_{\pm3.4307}$ & 0.9822$_{\pm0.0057}$ \\
$\Delta$-Diffusion~w/o~MD& 27.9003$_{\pm2.2482}$ & 0.8611$_{\pm0.0270}$ & 37.0520$_{\pm3.1037}$ & 0.9837$_{\pm0.0048}$ \\
%Bridge+Poisson+T1w  
$\Delta$-Diffusion~w/o~VD& 28.1462$_{\pm1.5705}$ & 0.8658$_{\pm0.0218}$ & 37.1123$_{\pm3.1751}$ & 0.9828$_{\pm0.0046}$ \\
%Bridge+Poisson+Delta d 
$\Delta$-Diffusion~w/o~MV& 28.6080$_{\pm2.2304}$ & \textbf{0.8746}$_{\pm0.0256}$ & 37.5737$_{\pm3.6653}$ & 0.9831$_{\pm0.0057}$ \\
\midrule
%Bridge+Poisson+T1w+Delta d+VOI loss 
$\Delta$-Diffusion & \textbf{28.6489}$_{\pm1.9933}$ & 0.8732$_{\pm0.0232}$ & \textbf{38.0669}$_{\pm3.5655}$ & \textbf{0.9841}$_{\pm0.0055}$ \\ 
\bottomrule
\end{tabular}
}
\label{tab_ablation}
\end{table}

\enlargethispage*{0.8\baselineskip}
\section{Conclusion}
This paper presents $\Delta$-Diffusion, a progression-aware diffusion framework for longitudinal synthesis of amyloid PET. 
By conditioning on MRI and elapsed follow-up interval, our model learns distributional transitions from baseline to follow-up PET via a Poisson Diffusion Bridge, rather than a baseline-conditioned noise-to-image generation process.
This approach suppresses trivial copying of baseline features and ensures both anatomical consistency and accurate temporal calibration.  
Extensive experiments on two cohorts with longitudinal PIB- and AV45-PET suggest that $\Delta$-Diffusion consistently outperforms SOTA methods.  

\section*{Acknowledgments}
This work was supported in part by NIH grants R01AG073297, RF1AG082938,
R01EB035160, and R01NS134849. 

%\newpage
\bibliographystyle{splncs04}
\bibliography{reference}

\end{document}